\documentclass[1p,12 pt, preprint,review]{article}
\usepackage{epsfig}
\usepackage{graphicx}
\usepackage{setspace}
\date{}
\def\b{\begin{equation}}
\def\e{\end{equation}}
\def\bee{\begin{enumerate}}
\def\eee{\end{enumerate}}
\begin{document}
\title{\bf Shadowing correction to the gluon distribution
behavior at small $x$}

\author{G.R.Boroun $^{1}$\thanks {Email:
grboroun@gmail.com}} \maketitle {\it \centerline{ \emph{
$^{1}$Physics Department, Razi University, Kermanshah 67149, Iran
}}
\newpage
\begin{abstract}
\emph{{We determined the saturation exponent of the gluon
distribution using the solution of the QCD nonlinear
Dokshitzer-Gribov-Lipatov-Altarelli-parisi (NLDGLAP) evolution
equation at small $x$. The very small $x$ behavior of the gluon
distribution is obtained by solving the Gribov, Levin, Ryskin,
Mueller and Qiu (GLR-MQ) evolution equation with the nonlinear
shadowing term incorporated. The form of initial condition for the
equation is determined. We find, with decreasing $x$, the
emergence of a singular behavior and the eventual taming (at
$R=5\hspace{0.1cm}GeV^{-1}$) and the essential taming (at
$R=2\hspace{0.1cm}GeV^{-1}$) of this singular behavior by
shadowing term. The nonlinear gluon density functions are
calculated and compared with the results for the integrated gluon
density from the Balitsky- Kovchegov(BK) equation for the
different values of $Q^{2}$. It is shown, that the results for the
gluon density function are comparable with the results obtained
from BK equation solution. Also we show that for each $x$, the
$Q^{2}$ dependence of the data is well described by gluon
shadowing corrections to GLR-MQ equation. The resulting analytic
expression allow us to predict the logarithmic derivative
$\frac{{\partial}F^{s}_{2}(x,Q^{2})}{{\partial}lnQ^{2}}$ and to
compare the results with H1 data and a QCD analysis fit. }}
\end{abstract}
 \vspace{0.5cm} {\it \emph{Keywords}}: \emph{shadowing
correction,nonlinear DGLAP evolution equation; GLR-MQ equation; Small-$x$ }\\
 \vspace{0.5cm} {\it \emph{PACS}}: \emph{
 13.85Hd, 12.38.Bx, 12.38.-t, 13.60.Hb}

\newpage
%%%%%%%%%%%%%%%%%%%%%%%%%%%%%%%%%%%%%%%%%%%%%%%%%%%%%%%%%%%%%%%%%
\emph{Deep inelastic electron- proton scattering (DIS) at small
value of the Bjorken variable $x$($<<$1) attracted a lot of
attention, mostly due to the experimental results from HERA. Of
great relevance is the determination of the gluon density at low
$x$, where gluons are expected to be dominant, because it could be
a test of perturbative quantum chromodynamic (PQCD) or a probe of
new effects, and also because it is the basic ingredient in many
other calculations of different high energy hadronic processes.}

\emph{In PQCD, the high- $Q^{2}$ behavior of DIS is given by the
Dokshitzer- Gribov- Lipatov- Altarelli- Parisi (DGLAP) evolution
equations [1]. In the double asymptotic limit (large energies,
i.e. small- $x$ and large photon virtualities $Q^{2}$), the DGLAP
evolution equations can be solved [2] and the structure function
is expected to rise approximately like a power of $x$ towards
small- $x$. This steep rise of $F_{2}(x,Q^{2})$ towards low- $x$
observed at HERA [3] also indicates in PQCD a similar rise of the
gluon towards low- $x$. This similar behavior predicts a steep
power law behavior for the gluon distribution. Accordingly, the
approximate solutions of DGLAP evolution equations have been
reported in recent years [4-6] with considerable phenomenological
success. In DIS at moderate values of $x$, the linear QCD
evolution equations lead to good description of this process. But
at small $x$, the problem is more complicated since recombination
processes between gluons in a dense system have to be taken into
account. This strong rise can eventually violate unitarity and so
it has to be tamed by screening effects. These screening effects
are provided by multiple gluon interaction which lead to the
nonlinear terms [7] in the DGLAP equations [8]. These nonlinear
terms reduce the growth of the gluon distribution in this
kinematic region where $\alpha_{s}$ is still small but the density
of partons becomes so large. Gribov, Levin, Ryskin, Mueller and
Qiu (GLRMQ)[7,9] performed a detailed study of this region. They
argued that the physical processes of interaction and
recombination of partons become important in the parton cascade at
a large value of the parton density, and that these shadowing
corrections could be expressed in a new evolution equation (the
GLRMQ equation). The main characteristic of this equation is that
it predicts a saturation of the gluon distribution at very small
$x$ [10]. This equation
was based on two processes in a parton cascade:\\
 i)The emission
induced by the QCD vertex $G{\rightarrow}G+G$ with the probability
which is proportional to $\alpha_{s}\rho$ where
$\rho(=\frac{xg(x,Q^{2})}{{\pi}R^{2}})$ is the density of gluon in
the transverse plane, ${\pi}R^{2}$ is the target area, and $R$ is the size of the target which the gluons populate;\\
ii)The annihilation of a gluon by the same vertex
$G+G{\rightarrow}G$ with the probability which is proportional to
$\frac{\alpha_{s}^{2}\rho^{2}}{Q^{2}}$, where $\alpha_{s}$ is
probability of the processes.}

\emph{When $x$ is not too small, only emission of new partons is
essential because $\rho$ is small and this emission is described
by the DGLAP evolution equations. Since gluon recombination
introduces a negative correction to the DGLAP evolution equations,
so the signal of its presence is a decrease of the scaling
violations as compared to the expectations obtained from the DGLAP
equations. However as $x\rightarrow0$ the value of $\rho$ becomes
so large that the annihilation of partons becomes important. This
picture allows us to write GLRMQ equation for the gluon structure
function behavior at small $x$, as[11],\b
\frac{dG(x,Q^{2})}{dlnQ^{2}}=\frac{\alpha_{s}}{2\pi}\int^{1-\chi}_{0}{dz}G(\frac{x}{1-z},Q^{2})P_{gg}(\frac{1}{1-z})
-\frac{81\alpha^{2}_{s}}{16R^{2}Q^{2}}\int^{1-\chi}_{0}\frac{dz}{1-z}[G(\frac{x}{1-z},Q^{2})]^{2},
\e Here the representation for the gluon distribution $G(x)=xg(x)$
is used, where $g(x)$ is the gluon density. As usual, $P_{gg}$ is
the gluon- gluon splitting function and $\chi=\frac{x}{x_{0}}$.
Where $x_{0}$ is the boundary condition that the gluon
distribution joints smoothly onto the unshadowed region. We
neglect the quark gluon emission diagrams due to their little
importance in the gluon rich low $x$ region.}

\emph{One of the striking discoveries at HERA is the steep rise of
the gluon distribution function with decreasing $x$ value
[3,6,12,13]. HERA shows that the gluon distribution function has a
steep behavior in the small $x$ region. Indeed, considering the
HERA data, shows by, $G(x,Q^{2})=A_{g}x^{-\lambda_{g}}$, where
$\lambda_{g}$ is the Pomeron intercept mines one. Clearly the
rapid increase of the gluon distribution with decreasing $x$
cannot go on indefinitely. When the density of gluons becomes too
large, they can no longer be treated as free partons. But the
shadowing corrections give rise to nonlinear terms (Eq.1) in the
evolution equation for $G(x,Q^{2})$. So, this singular behavior
tamed by these shadowing effects [14].} \emph{Strategy for
determination of the gluon distribution function with shadowing
correction also is based on the Regge- like behavior, as, \b
G^{s}(x,Q^{2})=A_{g}x^{-\lambda^{s}_{g}}.\e
 We note that at $x<x_{0}=10^{-2}$, shadowing and unshadowing gluon
distribution behavior are equal. At $Q^{2}_{0}$ the small $x$
behavior of the gluon distribution assumed to be[14],\b
G^{s}(x,Q^{2}_{0})=G^{u}(x,Q^{2}_{0})[1+\theta(x_{0}-x)
[G^{u}(x,Q^{2}_{0})-G^{u}(x_{0},Q^{2}_{0})]/xg_{sat}(x,Q^{2}_{0})]^{-1},\e
where, \b
xg_{sat}(x,Q^{2})=\frac{16R^{2}Q^{2}}{27\pi\alpha_{s}(Q^{2})},\e
is the value of the gluon which would saturate the unitarity limit
in the leading shadowing approximation. Based on this behavior,
the shadowing exponent of the gluon distribution can be determined
as, \b
\lambda^{s}(x,Q^{2}_{0})=\lambda^{u}(x,Q^{2}_{0})+\frac{1}{Ln{x}}
Ln[1+\theta(x_{0}-x)[G^{u}(x,Q^{2}_{0})-G^{u}(x_{0},Q^{2}_{0})]
\frac{27\pi^{2}}{4\beta_{0}R^{2}\Lambda^{2}t_{0}^{2}exp(t_{0})}]^{-1},\e
where
$\alpha^{LO}_{s}(Q^{2})=\frac{4\pi}{\beta_{0}\ln(\frac{Q^{2}}{\Lambda^{2}})}$,
$\beta_{0}=\frac{1}{3}(33-2N_{f})$ and $N_{f}$ being the number of
active quark flavors ($N_{f}=4$), also
t=ln$(\frac{Q^{2}}{\Lambda^{2}})$,
$t_{0}$=ln$(\frac{Q_{0}^{2}}{\Lambda^{2}})$ (that $\Lambda$ is the
QCD cut- off parameter). This equation (Eq.5) gives the shadowing
exponent of the shadowing gluon distribution function at the scale
$Q^{2}=Q^{2}_{0}$. In order to solve this equation we take
$\lambda^{u}_{g_{0}}$(=$\frac{{\partial}{\ln}G^{u}(x,t_{0})}{{\partial}{\ln}\frac{1}{x}}$)
that it is exponent at the starting scale $t_{0}$ while
$G^{u}(x,t_{0})$ is the input unshadowing gluon distribution that
take from GRV parametrisation [15]. The value of $R$ depends on
how the gluon ladders couple to the proton, or on how the gluons
are distributed within the proton. $R$ will be of the order of the
proton radius $(R\simeq5\hspace{0.1cm} GeV^{-1})$ if the gluons
are spread throughout the entire nucleon, or much smaller
$(R\simeq2\hspace{0.1cm} GeV^{-1})$ if gluons are concentrated in
hot- spot [16] within the proton.} \emph{In Fig.1, we show
$G^{s}(x,Q^{2})$ calculated from Eq.(5) as a function of $x$ at
scale $Q^{2}=5\hspace{0.1cm} GeV^{2}$ based on GRV unshadowing
gluon distribution parametrisation [15]. Input parametrisations
are obtained with help of GRV parametrisation of parton densities
and therefore could be used as independent parametrisation of
unintegrated gluon density. We observed that, as $x$ decreases,
the singularity behavior of the gluon function is tamed by
shadowing effects. These data show that gluon distribution
function increase as x decreases. They are corresponding with PQCD
fits at low x to data, but this behavior at low x tamed with
respect to nonlinear terms at GLR-MQ equation .}

\emph{ We perform the check of our calculations comparing
determined gluon density function with the solution of more
precise and more complicated  BK equation [17,18]. Where, the BK
equation was derived for the deep inelastic scattering of virtual
photon on a large nucleus by the resummation of multiple Pomeron
exchanges in the leading logarithmic approximation and in the
large $N_{c}$ limit. The BK equation is an evolution equation for
the imaginary part of the forward scattering amplitude of the
$q\overline{q}$ dipole on the nucleus. In the infinite momentum
frame, the BK equation resumes fan diagrams with the BFKL ladders,
corresponding to the Pomeron, splitting into two ladders. This
type of summation was originally proposed by GLR in the double
logarithmic approximation (DLA) ($\alpha_{s}<<1$ and
$\alpha_{s}ln(\frac{1}{x})ln(\frac{Q^{2}}{\Lambda^{2}}){\sim}1$).
The non-linear evolution equation for the amplitude reduce to the
GLR equation for the integrated gluon distribution. The model
based on nonlinear DGLAP shows a correspondence with BK equation
solution that calculate the values of integrated gluon
density[19,20,21] at the initial point. In order to evaluation
nonlinear DGLAP equation at different values of $Q^{2}$, let us
consider the shadowing DGLAP evolution equation (Eq.1) where we
have defined the shadowed gluon distribution function to be
according Eq.2. We then rewrite Eq.1 as,\b
\frac{dG^{s}}{dt}=\frac{3\alpha_{s}}{\pi}\int^{1-\chi}_{0}\frac{dz}{1-z}A_{g}(\frac{x}{1-z})^{-\lambda^{s}_{g}}
-\frac{81\alpha^{2}_{s}}{16R^{2}Q^{2}}\int^{1-\chi}_{0}\frac{dz}{1-z}[A_{g}(\frac{x}{1-z})^{-\lambda^{s}_{g}}]^{2}.\e
Integrating this equation over $z$ for $0<z<1-\frac{x}{x_{0}}$, we
find that,\b
\frac{dG^{s}}{dt}=\frac{3\alpha_{s}}{\pi}G^{s}\frac{1-\chi^{\lambda^{s}_{g}}}{\lambda^{s}_{g}}\\\nonumber
-\frac{81\alpha^{2}_{s}}{16R^{2}Q^{2}}(G^{s})^{2}\frac{1-\chi^{2\lambda^{s}_{g}}}{2\lambda^{s}_{g}}.
\e}

\emph{After successive differentiations of both sides of Eq.7, and
some rearranging, we find a nonlinear differential equation which
determines $G^{s}(x,t)$ in terms of $\lambda^{s}_{g}$ and
derivative of $\lambda^{s}_{g}$ with respect to $t$, as, \b
\frac{d\lambda^{s}_{g}}{dt}G^{s}(x,t)+\lambda^{s}_{g}\frac{dG^{s}(x,t)}{dt}=\frac{3\alpha_{s}}{\pi}G^{s}(x,t)
-\frac{81\alpha^{2}_{s}}{16R^{2}Q^{2}}(G^{s}(x,t))^{2}\e where we
used the exponent $\lambda^{s}_{g}$ as is given by, \b
\lambda^{s}_{g}=\frac{{\partial}{\ln}G^{s}(x,t)}{{\partial}{\ln}\frac{1}{x}}|_{t=constant}.\e
Next, we combine terms and define the shadowing gluon function by,
\b
\frac{d\lambda^{s}_{g}}{dt}=\frac{12}{\beta_{0}t}\chi^{\lambda^{s}_{g}}
-\frac{81\alpha^{2}_{s}}{32R^{2}Q^{2}}(1+\chi^{2\lambda^{s}_{g}})G^{s}.\e
We still need to solve Eqs.8 and 10 for the shadowing exponents,
as we obtain an inhomogeneous second-order differential equation
which determines $\lambda^{s}_{g}$ at any $t$ value. To simplify
the notation, we define $\eta$ by, \b
\eta=\frac{\frac{12}{\beta_{0}t}\chi^{\lambda^{s}_{g}}-\frac{d\lambda^{s}_{g}}{dt}}
{1+\chi^{2\lambda^{s}_{g}}}. \e
 The equation then reads, \b
\frac{d\lambda^{s}_{g}}{dt}\eta+\lambda^{s}_{g}\eta(\frac{2}{t}+1)+\lambda^{s}_{g}\frac{\frac{-12}
{\beta_{0}t^{2}}\chi^{\lambda^{s}_{g}}+\frac{12}{\beta_{0}t}\chi^{\lambda^{s}_{g}}
\ln{x}\frac{d\lambda^{s}_{g}}{dt}-\frac{d^{2}\lambda^{s}_{g}}{dt^{2}}}{1+\chi^{2\lambda^{s}_{g}}}
-\lambda^{s}_{g}\eta\frac{2\chi^{2\lambda^{s}_{g}}\ln{x}\frac{d\lambda^{s}_{g}}{dt}}{1+\chi^{2\lambda^{s}_{g}}}=\frac{12}{\beta_{0}t}\eta-2\eta^{2}
\e which defines the solution for $\lambda^{s}_{g}$ at any $t$
value with respect to $\lambda^{s}_{g_{0}}$. We have found that we
can parametrize the $\lambda^{s}_{g}$ determined as a second
degree polynomial in $lnQ^{2}$, in spite of that in the linear
DGLAP evolution equation this behavior for $\lambda_{g}$ is
linearly with respect to $lnQ^{2}$ only in a limited $lnQ^{2}$
interval. We emphasize at this point that the shadowing gluon
distribution function $G^{s}(x,t)$ is completely determined based
on $\lambda^{s}_{g}$ and its derivative with respect to t through
the expressions in Eq.10. The results are shown in Fig.2 and 3 for
$Q^{2}=20\hspace{0.1cm}GeV^{2}$ and $100\hspace{0.1cm}GeV^{2}$.
The top curve shows the small $x$ behavior of the gluon
distribution when shadowing is neglected, and the lower curves
show the effect of the shadowing contribution assuming first
$R=5\hspace{0.1cm}GeV^{-1}$ and, second, the more extreme "hot-
spot" example [22] with $R=2\hspace{0.1cm}GeV^{-1}$. We conclude
from Figs.1-3 that conventional shadowing
$(R=5\hspace{0.1cm}GeV^{-1})$ has little impact on the behavior of
the gluon in the region $10^{-4}{\leq}x$. The simple conclusions,
which could be obtained from the present plots, are the following.
First of all, the our results at hot- spot for each $Q^{2}$ give
 values comparable of the gluon density that are comparing with
integrated gluon density function which denote as BK curve. They
grow both with the rapidity $1/x$. Our data show that gluon
distribution function increase as x decreases, that its
corresponding with PQCD fits at low x, but this behavior tamed
with respect to nonlinear terms at GLR-MQ equation. The second
conclusion concerns the shape of the found curves. It can be
observed, that our curves at hot spot have a shapes similar to the
BK curve for gluon density function but with a completely
different slope. Also, these curves show that for $Q^{2}$ constant
value, there is a crossover point for both the curves where both
the predictions are numerically equal. It is easy to see, that
this cross point increase toward low $x$ as $Q^{2}$ increases.}

\emph{In order to comparing our results with the experimental
data, we follow  the role of shadowing correction to the evolution
of the singlet quark distribution with respect to GLR-MQ
equations. Indeed, the DGLAP evolution equation of the $Q^{2}$
logarithmic slope of $F_{2}$ at low $x$, in leading order, is
directly proportional to the gluon structure function. But,
according to these corrections (fusion of two gluon ladders),
evolution of the shadowing structure function with respect to
$lnQ^{2}$ is corresponding with modified DGLAP evolution equation
[22-25]. As we have, \b
\frac{{\partial}xq(x,Q^{2})}{{\partial}lnQ^{2}}=\frac{{\partial}xq(x,Q^{2})}{{\partial}lnQ^{2}}|_{DGLAP}-\frac{27\alpha_{s}^{2}}{160R^{2}Q^{2}}
[xg(x,Q^{2})]^{2},\e where the first term is the standard DGLAP
evolution equation. At low $x$, the nonsinglet contribution is
negligible and can be ignord, so that we find,\b
\frac{{\partial}F^{s}_{2}(x,Q^{2})}{{\partial}lnQ^{2}}=\frac{{\partial}F_{2}(x,Q^{2})}{{\partial}lnQ^{2}}|_{DGLAP}-\frac{5}{18}\frac{27\alpha_{s}^{2}}{160R^{2}Q^{2}}
[xg(x,Q^{2})]^{2}.\e } \emph{To find an analytical solution in
leading order, we note that the shadowing gluon  distribution
function has the Regee- like behavior corresponding to Eq.2.
Inserting Eq.2 in Eq.14, we can obtain the $Q^{2}$ shadowing
logarithms slope of $F_{2}$ at low $x$, as ,\b
\frac{{\partial}F^{s}_{2}(x,Q^{2})}{{\partial}lnQ^{2}}=\frac{5\alpha_{s}}{9\pi}T(\lambda_{g})G^{s}-\frac{5}{18}\frac{27\alpha_{s}^{2}}{160R^{2}Q^{2}}
[G^{s}]^{2},\e where,\b
T(\lambda_{g})=\int_{\chi}^{1}z^{\lambda^{s}_{g}}[z^{2}+(1-z)^2].\e
}
 \emph{We show a plot of $\frac{{\partial}F^{s}_{2}(x,Q^{2})}{{\partial}lnQ^{2}}$ in Fig.4 for a set of values of $Q^{2}$ at  $x$
 constant at hot spot point $R=2 GeV^{-1}$, compared to the values measured by the H1 collaboration
 [26] and a QCD fit based on ZEUS data [27]. From this figure one
 can see that GLR-MQ equation tamed behavior with respect to gluon
 saturation as $Q^{2}$ increases. Consequently, derivative of the
 shadowing structure function with respect to $lnQ^{2}$ at fixed
 $x$ is close to H1 data and QCD analysis fit. This shadowing
 correction suppress the rate of growth in comparison with the
 DGLAP approach.
 }

 \emph{To conclude, It is well known that at very low values
of $x$ the density of gluons in the proton target increases, and
eventually the unitarity bound is violated. This occurs even for
high values of $Q^{2}$ (small $\alpha_{s}$) where we would have
expected that PQCD should hold. At high density the recombination
of gluons becomes dominant, and must be included in the
calculations. Hence , the usual DGLAP evolution must be amended.
When the shadowing term is combined with DGLAP evolution in the
double leading log approximation (DLLA), then we obtain the GLR-MQ
equation for the integrated gluon. In the GLR-MQ approach, the
triple- Pomeron vertex is computed in the gluon scale. On the
other hand, this structure triple- Pomeron vertex extracted by BK
equation. Indeed these equations are related together, because BK
equation may be reduced to GLR-MQ form [28]. So that, we have
solved the DGLAP equation, with the nonlinear shadowing term
included, in order to determine the very small $x$ behavior of the
gluon distribution $G^{s}(x,Q^{2})$ of the proton. In this way we
were able to study the interplay of the singular behavior,
generated by the linear term of the equation, with the taming of
this behavior by the nonlinear shadowing term. With decreasing
$x$, we find that an ${\sim}x^{-\lambda^{s}_{g}}$ behavior of the
gluon function emerges from the GLR-MQ equation. The value of
$\lambda^{s}_{g}$ depends on the lower $\lambda^{s}_{g_{0}}$ and
is variable to evolution in $Q^{2}$. We compared our solutions,
with the gluon distribution obtained in a QCD analysis and BK
evolution equation. There should be no difference between our
results at $R=5\hspace{0.1cm}GeV^{-1}$ and the leading order
solutions for $x_{P}{\geq}x{\geq}10^{-3}$. The differences seen in
all figures between data at $R=5\hspace{0.1cm}GeV^{-1}$ and
$R=2\hspace{0.1cm}GeV^{-1}$ increases as $x$ decreases. We have
shown that also the singular behavior is confirmed at
$R=5\hspace{0.1cm}GeV^{-1}$ but this singular behavior tamed at
$R=2\hspace{0.1cm}GeV^{-1}$ with decreases $x$ as $Q^{2}$
increases. This leads to the essential conclusion that the
singularity of the shadowed gluon distribution function must be
controlled at hot-spot region. We note, that the obtained results
at hot spot are comparable with results obtained based on BK
evolution equation. It is important, because for our approach it
means that we not only reproduce the shadowed results for the
gluon density at hot spot point, but also that we concern the
shape of the found curve at hot spot is similar to BK curve and
there is a crossover point for the both curves at each $Q^{2}$
constat. Based on our present calculations we conclude that the
behavior of
$\frac{{\partial}F^{s}_{2}(x,Q^{2})}{{\partial}lnQ^{2}}$ as
measured at HERA, tamed based on gluon saturation at low $x$. Our
results show that the data can be described in PQCD taking into
account shadowing corrections. }

%%%%%%%%%%%%%%%%%%%%%%%%%%%%%%%%%%%%%%%%%%%%%%%%%%%%%%%%%%%%%%%%%%%%%%%%
\emph{
}
%%%%%%%%%%%%%%%%%%%%%%%%%%%%%%%%%%%%%%%%%%%%%%%%%%%%%%%%%%%

\newpage
\subsection*{Figure captions }
{\emph{Fig 1: The values of the $G^{s}(x,Q^{2})$ at
$Q^{2}=5\hspace{0.1cm} GeV^{2}$ from the solution of the
 initial condition shadowing effects based on Eq.3 with the boundary condition at $x_{0}=10^{-2}$.
  The  curve is the GRV QCD
 fit with shadowing neglected
   and data are our results with the shadowing term included with $R=5\hspace{0.1cm}GeV^{-1}$ (Up triangle) and
   $R=2\hspace{0.1cm}GeV^{-1}$ (Down triangle) that compared with BK integrated gluon density function
   (dot).\\\\
Fig 2 : The values of the shadowing gluon distribution function at
$Q^{2}=20\hspace{0.1cm} GeV^{2}$ by solving the GLR-MQ evolution
equation
  with the boundary condition at $x_{0}=10^{-2}$ and $Q_{0}^{2}=5\hspace{0.1cm} GeV^{2}$. The curve is the GRV QCD fit with shadowing neglected
   and our data show the effect of the conventional gluon shadowing with $R=5\hspace{0.1cm}GeV^{-1}$ (Up triangle) together
    with the more extreme "hot-spot" shadowing with $R=2\hspace{0.1cm}GeV^{-1}$ (Down triangle) that compared with BK integrated gluon density function (dot). The small different
   at $x_{0}=10^{-2}$ arises because the input $G^{s}(x,Q_{0}^{2})$
   to the GLR-MQ evolution equation at starting point at $Q_{0}^{2}=5\hspace{0.1cm}
   GeV^{2}$.\\\\
   Fig 3: The same as Fig.2 at
$Q^{2}=100\hspace{0.1cm} GeV^{2}$.\\\\
Fig 4: A plot of the derivative of the shadowing structure
function with respect to $lnQ^{2}$ vs. $Q^{2}$ for $x=0.0005$ (up
triangles), compared to data from H1 Collab.[26] (circles) with
total error, and also a QCD fit to ZEUS data that accompanied with
error bands associated with the parameter uncertainties [27](solid
lines).}

\newpage

\begin{figure}
\centering
  \includegraphics[width=15cm]{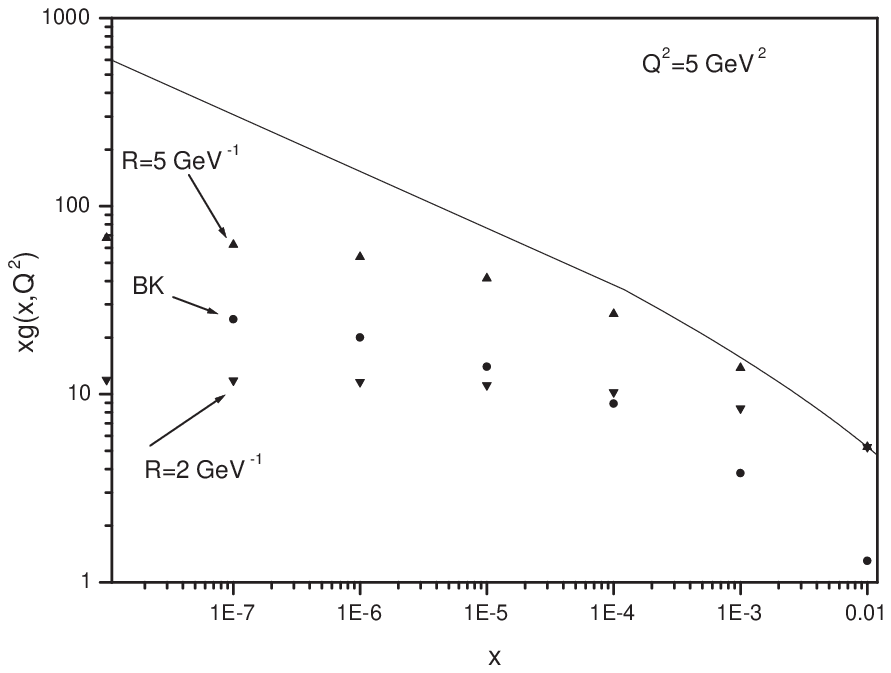}\\
  \caption{  }\label{}
\end{figure}

\begin{figure}
\centering
  \includegraphics[width=15cm]{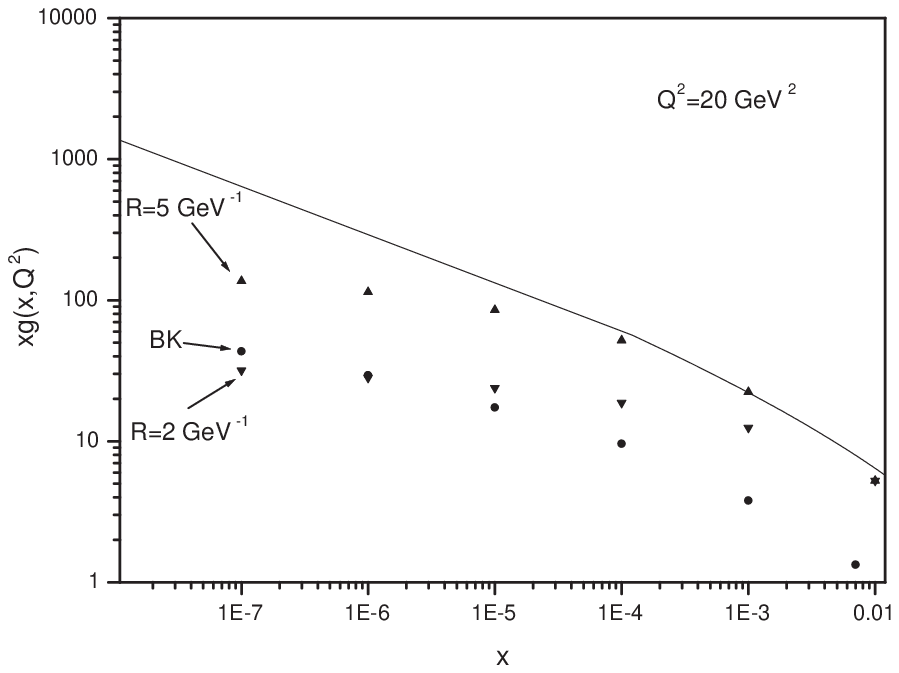}\\
  \caption{  }\label{}
\end{figure}
\begin{figure}
\centering
  \includegraphics[width=15cm]{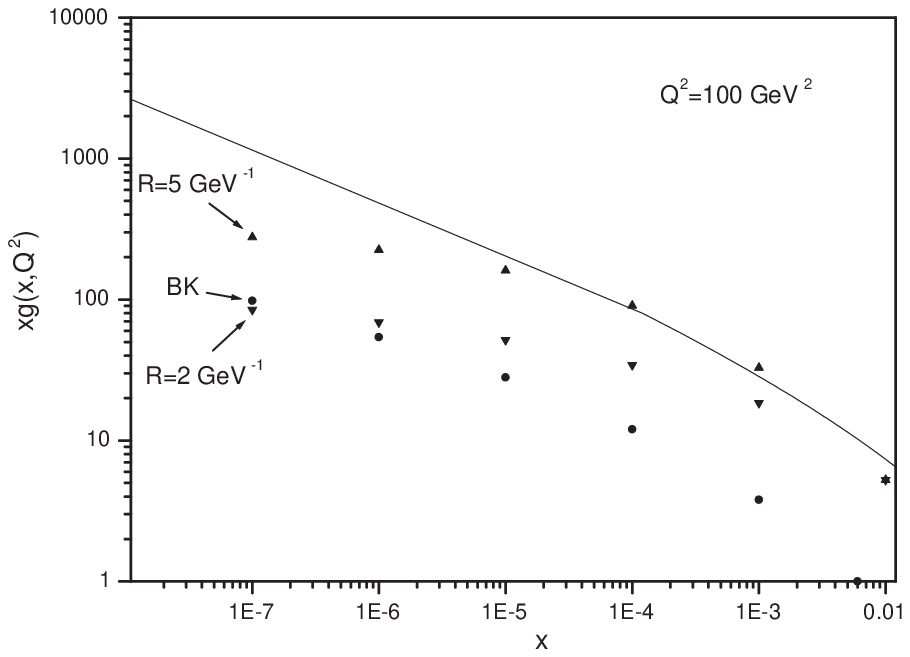}\\
  \caption{  }\label{}
\end{figure}
\begin{figure}
\centering
  \includegraphics[width=15cm]{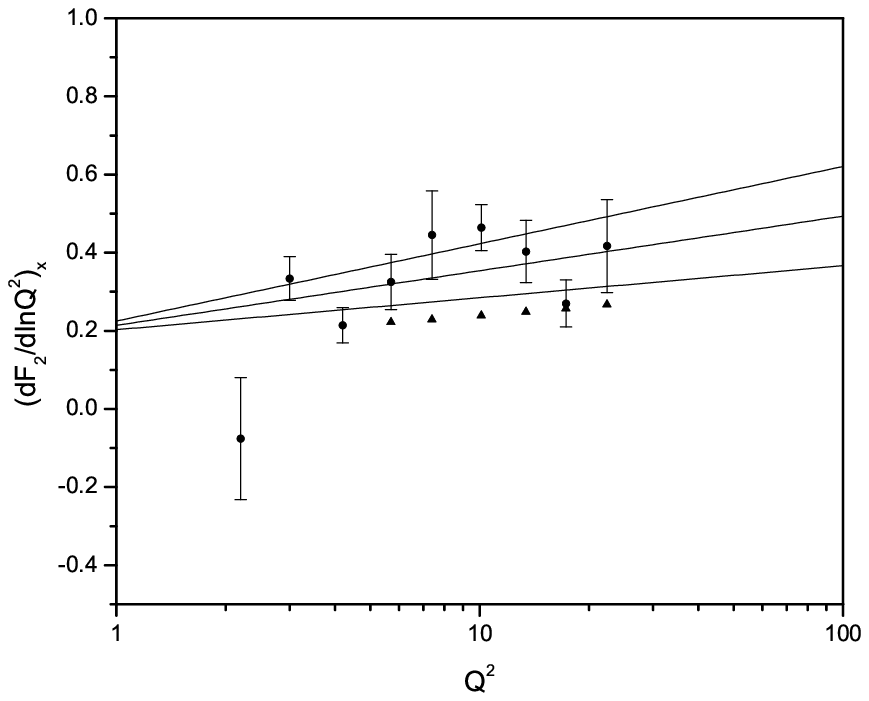}\\
  \caption{  }\label{}
\end{figure}
%%%%%%%%%%%%%%%%%%%%%%%%%%%%%%%%%%%%%%%%%%%%%%%%%%%%%%%%%%%%

\begin{thebibliography}{a}
\bibitem{h1}. Yu.L.Dokshitzer, Sov.Phys.JETP {\textbf{46}}, 641(1977);
G.Altarelli and G.Parisi, Nucl.Phys.B \textbf{126}, 298(1977);
V.N.Gribov and L.N.Lipatov, Sov.J.Nucl.Phys. \textbf{15},
438(1972).
 \bibitem{h2}. A.De Rujula \textit{et al}., phys.Rev.D \textbf{10}, 1649(1974);
 R.D.Ball and S.Forte, Phys.Lett.B \textbf{335}, 77(1994).
 \bibitem{h3}. $H1$ Collab., C.Adloff \textit{et al}., phys.Lett.B \textbf{520}, 183(2001).
 \bibitem{h4}. A.V.Kotikov and G.Parente, Phys.Lett.B \textbf{379}, 195(1996).
 \bibitem{h5}. J.K.Sarma and G.K.Medhi, Eur.Phys.J.C \textbf{16}, 481(2000).
 \bibitem{h6}.G.R.Boroun, JETP,\textbf{133}, No.4, 805(2008); G.R.Boroun and B.Rezaie, Phys.Atom.Nucl.\textbf{71}, No.6,
 1076(2008).
 \bibitem{h7}. A.H.Mueller and J.Qiu, Nucl.Phys.B\textbf{268}, 427(1986).
 \bibitem{h8}. J.Kwiecinski and A.M.Stasto, Phys.Rev.D\textbf{66}, 014013(2002).
 \bibitem{h9}. L.V.Gribov, E.M.Levin and M.G.Ryskin, Phys.Rep.\textbf{100},
 1(1983).
 \bibitem{h10}. A.L.A.Filho. M.B.Gay Ducati and V.P.Goncalves,
 Phys.Rev.D\textbf{59},054010(1999).
\bibitem{h11}. E.Laenen and E.Levin,Nucl.Phs.B.\textbf{451},207(1995).
\bibitem{h12}. A.M.Cooper- Sarkar and R.C.E.D Evenish, Acta.Phys.Polon.B \textbf{34}, 2911(2003).
\bibitem{h13}. A.Donnachie and P.V.Landshoff, Z.Phys.C \textbf{61},
139(1994); Phys.Lett.B \textbf{518}, 63(2001).
\bibitem{h14}. A.D.Martin, W.J.Stirling, R.G.Roberts,  and R.S.Thorne,
Phys.Rev.D \textbf{47}, 867(1993); J.Kwiecinski, A.D.Martin and
P.J.Sutton, Phys.Rev.D \textbf{44}, 2640(1991); A.J.Askew,
J.Kwiecinski, A.D.Martin and P.J.Sutton, Phys.Rev.D \textbf{47},
3775(1993).
\bibitem{h15}. M.Gluk, E.Reya and A.Vogt,Z.Phys.C \textbf{67}, 433(1995);
 Eur.Phys.J.C \textbf{5}, 461(1998).
\bibitem{h16}. E.M.Levin and M.G.Ryskin, Phys.Rep.\textbf{189}, 267(1990).
\bibitem{h17}. I.I.Balitsky, Nucl.Phys.B\textbf{463}, 99(1996).
\bibitem{h18}. Yu.V.Kovchegov, Phys.Rev.D\textbf{60},
034008(1999) $\&$ Phys.Rev.D\textbf{61}, 074018(2000).
 \bibitem{h19}. K.Golec-Biernat, L.Motyka, and A.M.Stasto, Phys.Rev.D\textbf{65},
074037(2002).
 \bibitem{h20}. J.Kwiecinski and A.M.stasto,Phys.Rev.D\textbf{66},
014013(2002).
 \bibitem{h21}. S.Bondarenko, Phys.Lett.B\textbf{665},
72(2008).
 \bibitem{h22}. A.H. Mueller,arXiv:hep-ph/0111244(2001).
 \bibitem{h23}. E.Gotsman,et.al., Nucl.Phys.B\textbf{539},
535(1999).
 \bibitem{h24}. K.J.Eskola,et.al., Nucl.Phys.B\textbf{660},
211(2003).
 \bibitem{h25}. E.Gotsman,et.al., Phys.Lett.B\textbf{500},
87(2001).
\bibitem{h26}. $H1$ Collab., C.Adloff \textit{et al}., Eur.Phys.J.C \textbf{13},
609(2000); Eur.Phys.J.C \textbf{21}, 33(2001).
\bibitem{h27}. E.L.Berger, M.M.Block and C.I Tan, Phys.Rev.Lett\textbf{98},
242001(2007).
\bibitem{h28}. M.A.Kimber, J.Kwiecinski and A.D.Martin,
Phys.Lett.B508,58(2001).
\end{thebibliography}
\end{document}